\documentclass[]{aa}
\usepackage{graphicx}
\usepackage{natbib}
\usepackage{txfonts}
\def\mh{\,$\mu$Hz}
\def\fm{$\nu_\mathrm{max}$}
\def\dnu{$\Delta \nu$}
\def\lg{\ensuremath{\log g}}
 
\begin{document}
   \title{Oscillating red giants in the CoRoT\thanks{The CoRoT ({\it \underline{Co}nvection, \underline{Ro}tation, and planetary \underline{T}ransits}) space mission, launched on 2006 December 27, was developed and is operated by the CNES, with participation of the Science Programs of ESA, ESAs RSSD, Austria, Belgium, Brazil, Germany and Spain.} exofield: Asteroseismic mass and radius determination}

   \author{T. Kallinger\inst{1, 2}
   	 \and
         W. W. Weiss\inst{2}
         \and
         C. Barban\inst{3}
         \and
         F. Baudin\inst{3}
         \and
         C. Cameron\inst{4}
         \and
         F. Carrier\inst{5}
         \and
         J. De Ridder\inst{5}
         \and
         M.-J Goupil\inst{3}         
         \and
         \\M. Gruberbauer\inst{4, 2}
         \and
         A. Hatzes\inst{6}
         \and 
         S. Hekker\inst{7, 8, 5}
         \and
         R. Samadi\inst{3}
         \and 
         M. Deleuil\inst{9}
          }

   \offprints{kallinger@phas.ubc.ca}

   \institute{Department of Physics and Astronomy, University of British Columbia, 6224 Agricultural Road, Vancouver, BC V6T 1Z1, Canada
		\and 
Institute for Astronomy, University of Vienna, T\"urkenschanzstrasse 17, 1180 Vienna, Austria
		   \and	
Observatoire de Paris, LESIA, CNRS UMR 8109 Place Jules Janssen, F-92195 Meudon, France
              \and
Department of Astronomy and Physics, Saint Mary's University, Halifax, NS, B3H 3C3, Canada
              \and		   
Instituut voor Sterrenkunde, Katholieke Universiteit Leuven, Celestijnenlaan 200 B, 3001 Heverlee, Belgium
		   \and
Th\"uringer Landessternwarte Tautenburg, Sternwarte 5, 07778 Tautenburg, Germany
		   \and		   
University of Birmingham, School of Physics and Astronomy, Edgbaston, Birmingham B15 2TT, United Kingdom
	\and
Royal Observatory of Belgium, Ringlaan 3, 1180 Brussels, Belgium
		\and
Laboratoire dAstrophysique de Marseille (UMR 6110), Technople de Marseille-Etoile, F-13388 Marseille cedex 13, France
             }

   \date{Received 2008 November 28; accepted 2009 November 17}

\abstract
{Observations and analysis of solar-type oscillations in red-giant stars is an emerging aspect of asteroseismic analysis with a number of open questions yet to be explored. 
Although stochastic oscillations have previously been detected in red giants from both radial velocity and photometric measurements, those data were either too short or had sampling that was not complete enough to perform a detailed data analysis of the variability. The quality and quantity of photometric data as provided by the CoRoT satellite is necessary to provide a breakthrough in observing p-mode oscillations in red giants. 
We have analyzed continuous photometric time-series of about 11\,400 relatively faint stars obtained in the \emph{exofield} of CoRoT during the first 150 days long-run campaign from May to October 2007. We find several hundred stars showing a clear power excess in a frequency and amplitude range  expected for red-giant pulsators. In this paper we present first results on a sub-sample of these stars.}
{Knowing reliable fundamental parameters like mass and radius is essential for detailed asteroseismic studies of red-giant stars. As the CoRoT exofield targets are relatively faint (11--16\,mag) there are no (or only weak) constraints on the stars' location in the H-R diagram. We therefore aim to extract information about such fundamental parameters solely from the available time series.}
{We model the convective background noise and the power excess hump due to pulsation with a global model fit and deduce reliable estimates for the stellar mass and radius from scaling relations for the frequency of maximum oscillation power and the characteristic frequency separation.}
{We provide a simple method to estimate stellar masses and radii for stars exhibiting solar-type oscillations. Our method is tested on a number of known solar-type pulsators.}
{}

   \keywords{stars: oscillations -- stars: fundamental parameters -- techniques: photometric}
\authorrunning{Kallinger et al.}
\titlerunning{Oscillating red giants in the CoRoT exo-field}
   \maketitle

\section{Introduction}

Stars cooler than the red border of the instability strip have convective envelopes where turbulent motions act over various time scales and velocities (up to the local speed of sound), producing acoustic noise which can stochastically drive (or damp) resonant, p-mode oscillations. All cool stars with convective outer layers potentially show these solar-type oscillations typically with small amplitudes. The oscillation amplitudes are believed to scale with the luminosity and are, therefore, more easily observed in evolved red giants than in main-sequence stars, which opens up a promising potential for asteroseismic investigations of evolved stars. Their larger radii, however, adjust the pulsation periods from minutes to several hours to days. This in turn complicates ground-based detection and calls for long and uninterrupted observations from space. 

It is believed that the global characteristics of solar-type oscillations, like the frequency range of pulsation or their amplitudes, are predetermined by the global properties of the star, like its mass or radius. It should therefore be possible to deduce the stellar fundamental parameters of a solar-type pulsator from the global properties of the observed oscillations. Recent investigations in this context were made by \citet{gil08} ,who analyzed pulsation amplitudes and timescales in several hundred red giants in the galactic bulge observed by the Hubble Space Telescope and \citet{ste08}, who determined asteroseismic masses for eleven bright red giants observed with the star tracker of the WIRE satellite. In this paper we measure global asteroseismic quantities for 31 red giants observed with the CoRoT satellite and use well-known scaling relations to estimate their masses and radii.

The satellite CoRoT \citep{bag06} is continuously collecting three-color photometry for thousands of relatively faint (about 11--16\,mag) stars in the so-called \emph{exofield} with the primary goal to detect planetary transits. But the data are also perfectly suited for asteroseismic investigations. And indeed, a first processing of the CoRoT exofield data reveals a variety of oscillating red giants. \citet{hek09} report on the clear detection of solar-type oscillations in several hundreds of red giants among the $\sim$11\,400 stars observed during the first 150 days CoRoT long-run (LRc01) campaign.

First results for oscillations in red giants observed in the CoRoT exofield are presented in \citet{rid09}, who also discuss the question of the existence of non-radial modes in red giants with moderate mode lifetimes versus the presence of short living radial modes only. Briefly, \citet{bar07} interpreted the signal found in the MOST observations of the red giant $\epsilon$ Oph as radial modes with relatively broad profiles corresponding to short mode lifetimes of $\sim$2.7 days. Their result was consistent with what was found for the similar red giant $\xi$ Hya \citep{ste06}. On the other hand, \citet{kal08} re-examined the same data set and found radial \emph{and} non-radial modes with significantly longer mode lifetimes (10--20 days), which supports the lifetimes inferred from theoretical considerations by \citet{hou02} and only recently \citet{dup09}. Additionally, the frequencies are consistent with those of a red-giant model that matches $\epsilon$ Ophs' position in the H-R diagram. So the actual interpretation of the available observations was unclear. It was \citet{rid09} who first provided unambiguous evidence for regular p-mode patterns of radial and non-radial modes with long lifetimes.

The low-degree, high-radial order p modes of solar-type pulsators are approximately equally spaced and are believed to follow an asymptotic relation \citep{tas80}. For a detailed asteroseismic analysis, such as the comparison of the individual observed frequencies and/or their separations with those of stellar models, is it often important to constrain the parameter space by using reliable fundamental parameters. Apart from pulsation, solar-type pulsators also show significant power from the turbulent fluctuations in their convective envelopes opening the possibility to study convective time scales and amplitudes. Without knowing the position of the star in the H-R diagram such endeavours cannot be properly realized.

This knowledge is indeed lacking for most of the faint CoRoT exofield stars. The available broadband color-information allows at best, and only in some cases, a rough determination of the effective temperature, but is not suitable to distinguish giants from main-sequence stars. We therefore try to extract the fundamental parameters from the time series alone, which is the main topic of this paper.
We model the convective background noise and the power excess due to pulsation with a global fit, which allows us to measure the so-called frequency of maximum oscillation power. With this and the large frequency separation we derive the stellar mass and radius from well-known scaling relations. As a first estimate we also obtain effective temperatures and luminosities from a comparison with evolutionary tracks.

   \begin{figure}[t]
   \centering
      \includegraphics[width=0.5\textwidth]{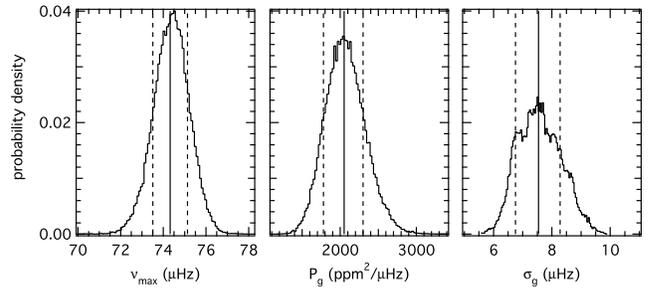}
      \caption{Marginal distributions of the Gaussian parameters (Eq.\,\ref{eq:fit}) for the red giant  ``A" as computed by the MCMC algorithm. Median values and 1$\sigma$ limits are indicated by vertical solid and dashed lines, respectively.}
         \label{Fig:Distri}
   \end{figure}

   \begin{figure}[b]
   \centering
      \includegraphics[width=0.5\textwidth]{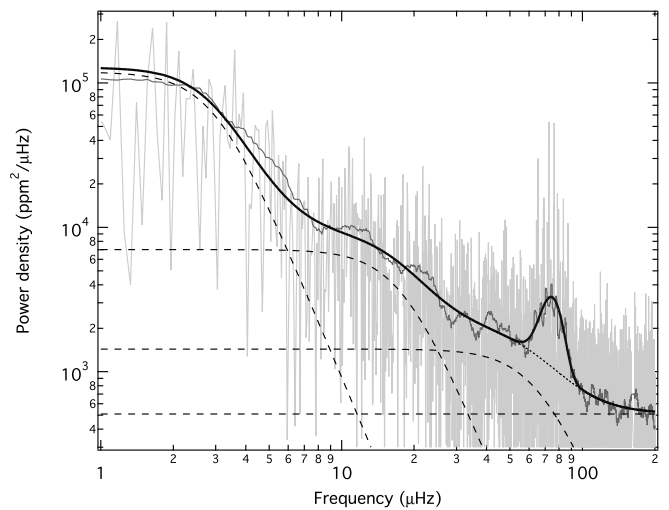}
      \caption{Original (light-grey) and heavily smoothed (dark-grey) power density spectrum of the CoRoT exofield time series of the red giant A and a global model (black line) fitted to the power density spectrum. The model is a superposition of white noise (horizontal dashed line), three power law components (dashed lines) and a power excess hump approximated by a Gaussian function. The dotted line indicates the model fit plotted without the Gaussian component.}
         \label{Fig:Pow}
   \end{figure}


   \begin{figure*}[t]
   \centering
      \includegraphics[width=1.02\textwidth]{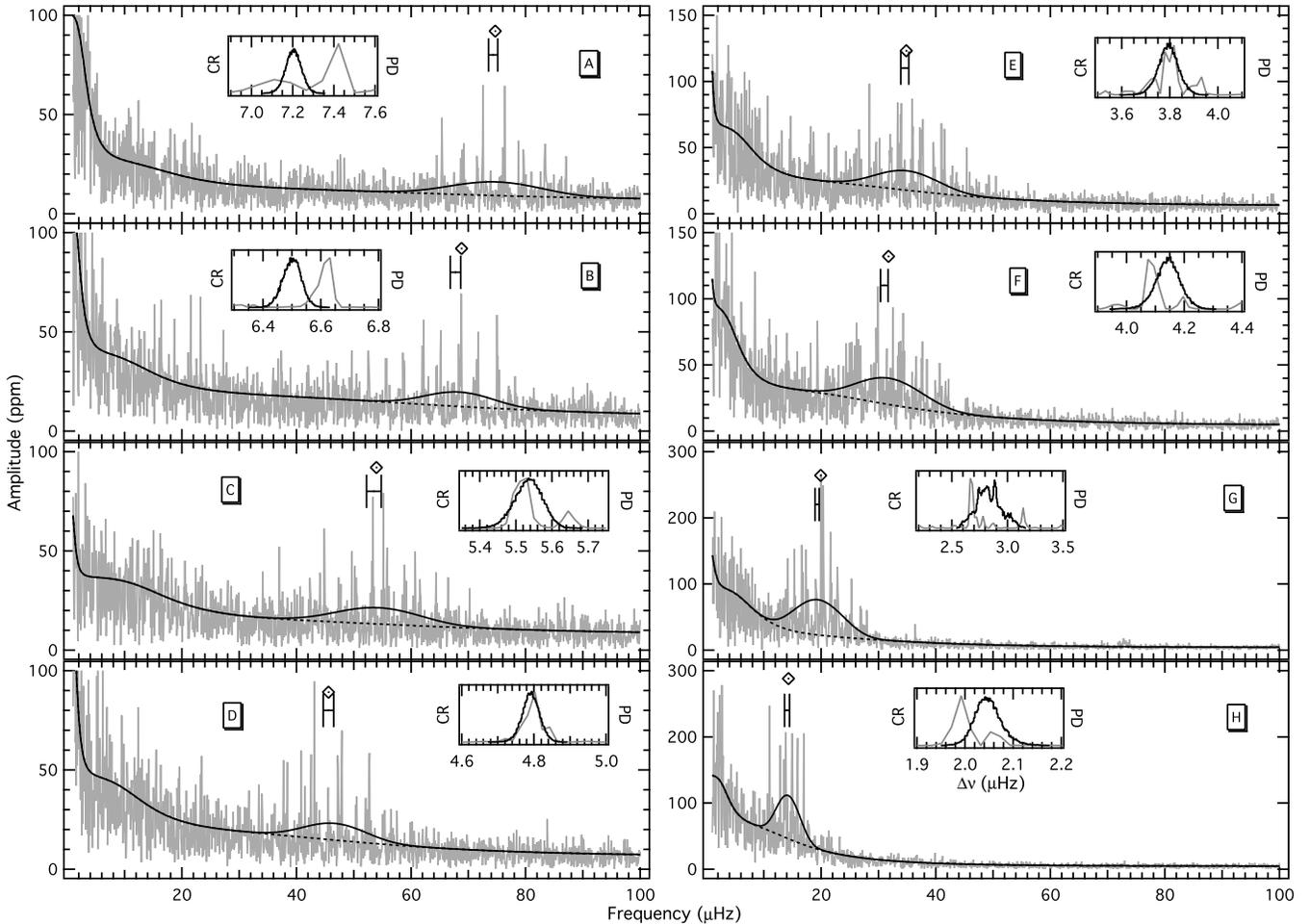}
      \caption{Fourier amplitude spectra for a sample of red giants observed by CoRoT. Black lines indicate a global model fit, and dotted lines show the model plotted without the Gaussian component and serve as a model for the background signal. The center of the Gaussian is adopted to be the frequency of maximum oscillation power, $\nu_\mathrm{max}$. Error bars indicate the $\pm$1$\sigma$ uncertainties and diamond symbols correspond to the weighted mean frequency after correcting for the background signal. Note the different amplitude scales in the different panels. Inserts compare the comb response functions (grey) and marginal distributions (black) used to determine the large frequency separations.}
         \label{Fig:Fourier}
   \end{figure*}

\section{Observations}
Although the CoRoT satellite and the exofield data are well explained by others \citep[][and references therein]{rid09}, we will briefly summarize the instrument and data sets.

CoRoT houses four 1k $\times$ 1k pixels CCD photometers fed by a 27-cm afocal telescope. The satellite's low-Earth polar orbit (period $\approx$ 100\,min; $\approx$ 167\mh ) enables uninterrupted observations of stars in its continuous viewing zones (two cones with $\sim$10\degr\ radius centered on the galactic plane at right ascension of about 6:50\,h and 18:50\,h, respectively) for up to six months. A summary of the mission is given in the pre-launch proceedings of CoRoT published by ESA (SP-1306, 2006). The two core science objectives, asteroseismology across the H-R diagram and the detection of transiting extra-solar planets, are tracked simultaneously with two of the four detectors each. In order to check the color independency of presumed planetary transits, the stellar light is dispersed by a prism before it reaches the exofield detectors. In this paper we concentrate on the white light flux measurements which are obtained by adding the flux of the three color channels to improve the photometric quality.

During the first 150 day long-run campaign, CoRoT pointed toward the coordinates ($\alpha, \delta$) = (19.4\,h , 0.46\degr ) from May to October, 2007, and gathered time series of about 11\,400 stars sampled with a cadence of either 512\,s or 32\,s, depending on the predefined status of the star (the limited downlink capacity does not allow to sample all stars with the short cadence). Typically, each time series consists of about 25\,000 or rather 400\,000 data points with a duty cycle of more than 90\%. We use the N2 data format \citep{sam07}, which is the output of a standard data reduction procedure, and detect and remove occasional jumps in the time series (caused by high energy particles) and apply an outlier correction. In a next step, we compute Fourier power spectra of all time series and extract parameters which we believed to be characteristic for red giant pulsators, like the 1/f$^{2}$ characteristic or the existence of a power excess hump. Based on a pre-selection with these parameters, we use a semi-automatic routine to identify the pulsating red giants. A more detailed description of how to identify the red giant stars is given in \citet{hek09}. The data are available at the CoRoT download page\footnote{http://idoc-corot.ias.u-psud.fr/}.

\section{Power spectra modeling}
The turbulent motions in the convective envelopes of cool stars act on a similar time scale as the acoustic oscillations and potentially complicate the detection and analysis of solar-type oscillations. Although the convective signal is stochastic, it follows particular characteristics. It can be shown that such quasi-stochastic variations cause correlations of consecutive measurements with the strength of the correlations exponentially decreasing for increasing time-lags. The Fourier transform of this correlated colored ``noise" follows a power law, characterized by an amplitude and a characteristic frequency (or inverse time scale). 

For the Sun, it is common practice to model the background signal with power laws to allow accurate measurements of the solar oscillation parameters. 
Power law models were first introduced by \citet{har85}. \citet{aig04} and only recently \citet{mic08} use the sum of power laws:  $P(\nu)=\sum_i P_i$, with $P_i=a_i \zeta_i^2 \tau_i / (1+ (2\pi \tau_i \nu)^{C_i})$ to fit the solar background, with $\nu$ being the frequency, $\tau_i$ the characteristic time scale, and $C_i$ the slope of the power law. $a_i$ serves as normalization factor for $\zeta_i^2 = \int P_i(\nu) d\nu$, which corresponds to the variance of the stochastic variation in the time domain. The slope of the power laws was originally fixed to 2 in Harvey's models, but \citet{aig04} and \citet{mic08} have shown that, at least for the Sun, the slope is closer to 4. The number of power law components usually varies from two to five, depending on the frequency coverage of the observations. Each power law component is believed to represent a different class of physical processes such as stellar activity, activity of the photospheric/chromospheric magnetic network, or granulation (see \citealt{aig04} or \citealt{mic08} and references therein) with time scales for the Sun ranging from months for active regions to minutes for granulation.  


First tests with power law fits to the CoRoT photometry have shown that the presence of an additional power due to pulsations significantly distorts such a fit and requires an additional component to model the entire spectrum. Since the shape of the pulsation power excess seems to be well approximated by a Gaussian, we model the observed power density spectra with a superposition of white noise, the sum of power laws, and a power excess hump approximated by a Gaussian function

\begin{equation}
P(\nu) =  P_n +  \sum_{i} \frac{A_{i}}{1 + (\nu / B_i)^4} + P_g \cdot e^{\,\,-(\nu_\mathrm{max} - \nu)^2 / (2\sigma_g^2)},
\label{eq:fit}
\end{equation}
where $P_n$ represents the white noise contribution and $A_{i}$ and $B_{i}$ are the amplitudes of the stellar background components and their characteristic frequencies (or inverse time scales), respectively. The frequency coverage which results from the 150\,d CoRoT observations is sufficient to use three power law components. $P_g$, $\nu_\mathrm{max}$, and $\sigma_g$ are the height, the central frequency, and the width of the power excess hump, respectively. 

We use a Bayesian Markov-Chain Monte Carlo (MCMC) algorithm to fit the global model to the power density spectrum. The algorithm samples a wide parameter space and delivers probability distributions for all relevant quantities. The procedure is described in \citet{gru08} and was originally designed to fit Lorentzian profiles to the p-mode spectrum of a solar-type pulsator. Our problem is quite similar, and it was trivial to adapt the code in order to fit a global model to the power density spectra instead of a sequence of mode profiles. The advantage of the algorithm is its stability and insensitivity to wrong initial parameters, and also that it delivers reliable parameters as well as realistic uncertainties in a fully automatic way. For the frequency parameters ($B_i$) we have sampled the entire frequency range of interest from 0 to 150\mh . For higher frequencies, the power spectra are potentially contaminated by instrumental artifacts due to the satellites' orbital period. The amplitude parameters ($P_n$, $A_i$, and $P_g$) were allowed to vary from zero to the highest amplitude peak in the spectrum. Only $\nu_{max}$ and $\sigma_g$ were kept within reasonable limits (0.5 to 2 times the value we inferred from a visual inspection of the spectra). After some 500\,000 iterations we calculated the most probable value and its 1$\sigma$ uncertainty for all fitted parameters from their marginal distribution and constructed the most probable global model fit.  

As an example, we show in Fig.\,\ref{Fig:Distri} the marginal distributions of the Gaussian parameters as computed by the MCMC algorithm for star A. The corresponding most probable global fit is given in Fig.\,\ref{Fig:Pow} along with the original and heavily smoothed power density spectrum, the white noise, and power law components of the fit. Fig.\,\ref{Fig:Fourier} shows a sequence of amplitude spectra selected from the 31 analyzed red giants with the corresponding global model fits. Note that the fits are calculated in power but are presented in amplitude for better visibility. The sequence impressively demonstrates that the frequency and amplitude range of the oscillations scale with the time scales and amplitudes of the background signal. 

In a next step we use the white noise and power law components of the global model (doted lines in Fig.\,\ref{Fig:Pow} and \ref{Fig:Fourier}) to correct the power density spectra for the background signal, which should then include only the oscillation signal. But what is more interesting in this context, we assume the center of the Gaussian to locate the centroid of the power excess hump, and as long as the power excess hump is symmetric, the center of a Gaussian fit should equal the frequency of maximum oscillation power. 
To test this assumption we compute the weighted mean frequency, $\bar{\nu}$, in the frequency range of pulsations (\fm\ $\pm 5\sigma_g$), where we use the residual power after correcting for the background signal as weight. Although for seven of the eight stars shown in Fig.\,\ref{Fig:Fourier}, $\bar{\nu}$ is within \fm $\pm$1$\sigma$, $\bar{\nu}$ seems to be systematically shifted towards higher frequencies. Indeed, for 19 out of the 31 stars in our sample, $\bar{\nu}$ is higher than \fm . But as the average shift for the 19 stars (0.6\%) is about four times smaller than the stochastic error of \fm , we expect the systematic error on our estimate of \fm\ to be negligible. Note that for the Sun the effect can not be neglected. Based on SOHO/VIRGO data \citep{fro97}, we find $\bar{\nu}$ to be shifted by almost 2\% towards higher frequencies.

The second interesting parameter that can directly be determined from the observed power spectrum is the large frequency separation, \dnu , of consecutive radial overtone modes of the same spherical degree. Since this frequency separation is, at least for the Sun \citep[see, e.g.,][]{bro09}, a function of the frequency itself, it is not straightforward to identify an average value for all observed modes. Such a value depends on the actual number and frequency range of the observed modes and is difficult to compare for different observations. We therefore specifically chose to define \dnu\ as the average frequency separation in the frequency range of the maximum oscillation power.
To identify \dnu\ we use again the Bayesian MCMC algorithm \citep{gru08}. But instead of fitting a global model, we fit a sequence of equidistant Lorentzian profiles to the power density spectra
\begin{equation}
P(\nu) =  P_n +  \sum_{i=-2}^{2} \frac{a_{i}^2 \cdot \tau}{1 + 4[\nu - (\nu_0 +\Delta\nu \cdot i/2)]^2 \cdot (\pi \tau)^2},
\label{eq:fit}
\end{equation}
where $a_i$ is the rms mode amplitude of the $i$-th profile, $\nu_0$ corresponds to the frequency of the central mode, \dnu\ is the spacing, and $\tau$ is the mode lifetime which is assumed to be equal for all five modes. In principle we do not expect all modes to have the same lifetime, but this is a marginal assumption which significantly stabilizes the fit and has very little impact on the determination of \dnu . The model obviously represents either three consecutive radial modes with two interjacent dipole modes or three consecutive dipole modes with interjacent radial modes, depending on which mode the sequence is centered. In either case, the fitted spacing corresponds to the large frequency separation in the frequency range where the maximum oscillation power is seen. In our analysis the MCMC algorithm is allowed to vary the central mode frequency within \fm $\pm\sigma_g$. \dnu\ and the mode lifetime are sampled between 0.5\mh\ and 2$\sigma_g$ and 1 and 100 days, respectively. The individual mode amplitudes are allowed to vary between zero and four times the highest amplitude peak in the spectrum. After half a million iterations the algorithm delivers probability distributions for all fitted parameters and we calculate the most probable values and their uncertainties from their marginal distributions. 

\begin{figure}[t]
   \centering
      \includegraphics[width=0.5\textwidth]{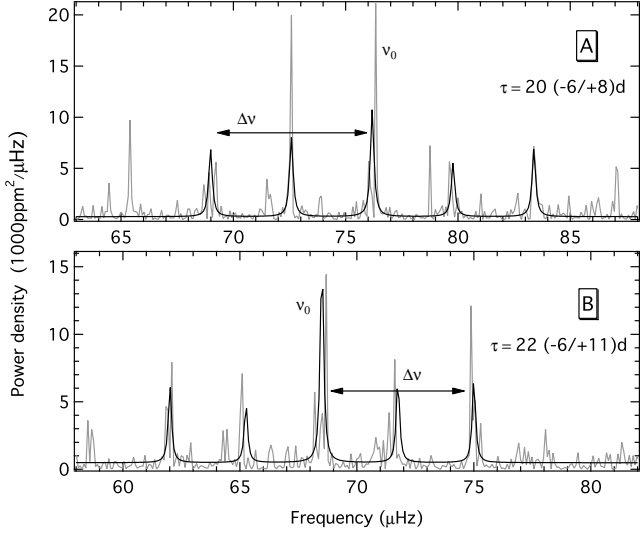}
      \caption{Residual power density spectra for star A and B after correcting for the background signal.  The most probable model-fits used to determine \dnu\ are indicated by black lines. }
         \label{Fig:lorfit}
   \end{figure}

The advantage of our method compared to, e.g., the comb-response function \citep{kje95a} or an autocorrelation spectrum is that it takes the Lorentzian-like form of the signal into account and is therefore less sensitive to the stochastic nature of the signal. Examples for the residual power density spectra and the most probable fits are shown in Fig.\,\ref{Fig:lorfit}. Interestingly, the presence of additional modes which are not taken into account in our model does not influence the fit. This can be seen for instance from the power density spectrum of star A, where the MCMC algorithm correctly identifies the $l$ = 0 and 1 modes and does not consider the additional peaks at about 72 and 79\mh , which are most likely $l$ = 2 modes. We compare the marginal distributions for \dnu\ from our MCMC algorithm with the comb-response functions (both with arbitrary ordinates) in the inserts of Fig.\,\ref{Fig:Fourier}. Although for some stars both methods give consistent results, the values can differ by more that 0.2\mh .

We expect this ambiguity to be due to the stochastic nature of the signal. The observed time series represent a single realisation of a damped and stochastically excited signal. Another realisation might result in a different measurement for \dnu\ and \fm . In order to examine this we simulate different realisations of the same solar-type oscillation signal. We use the original time series of star A and pre-withen all significant peaks in the frequency range of pulsation. The residual time series should now include only the intrinsic background signal. We then generate 250 time series following the procedure in \citet{cha97} with each data set representing a different realisation of four radial orders of equidistant $l$ = 0, 1, and 2 modes with \dnu\ = 7.20\mh\ and an arbitrary value for $\delta\nu$ = $\nu_{n,l} - \nu_{n-1,l+2}$ = 1\mh . The time domain rms amplitudes of radial and $l$ = 1 modes are set in a way that the mode heights follow the Gaussian shape of the original power excess hump. The rms amplitudes of $l$ = 2 modes are set arbitrarily to half the value of the closest radial mode. The largest amplitude mode is centered on 76.2\mh , and all modes have a lifetime of 20 days, which is a typical value for intermediate luminous red giants \citep{dup09} and corresponds to what we have found for the intrinsic modes. 
Finally we superpose each simulated data set with the residual time series, calculate the power density spectrum, and apply our algorithms to determine \fm\ and \dnu . The simulations give an average value \fm\ = 74.42\mh\ with a rms scatter of 0.55\mh ,  which is well within the 1$\sigma$ uncertainty of the originally determined value (74.32$\pm$0.81\mh ). The situation is different for \dnu . If we determine \dnu\ from the frequency of the largest peak of the comb-response function, then the stochastic nature of the signal adds a rms scatter of about 0.21\mh\ in our simulation. This is much compared to the rms scatter of only 0.04\mh\ if we use our Bayesian MCMC approach, which is again compatible with the 1$\sigma$ uncertainty of the original value (0.04\mh ). This assures us that the stochastic nature of the oscillation signal adds no significant additional uncertainty in our subsequent analysis. We list \fm , \dnu\ and the corresponding 1$\sigma$ uncertainties for our sample of red giants in Table\,\ref{tab:Cstars}.

\section{Asteroseismic determination of fundamental parameters}

Stellar masses of field stars are usually determined by comparing the location in the H-R diagram with evolutionary tracks. But when stars evolve to become red giants, their evolutionary tracks move together and a relatively narrow range in the H-R diagram covers a large range in mass. Consequently, the mass determined from a location in the H-R diagram becomes quite uncertain. Additionally, the actual position and slope of the red-giant branch in stellar evolutionary calculations depend very much on the parameters used to compute the models. It is, however, believed that the global properties of solar-type oscillations, like the frequency range where they can be observed, depend on the fundamental parameters of the star. It should therefore be possible to determine these parameters from global properties of the observed oscillations.

The amplitudes of solar oscillations are modulated by a broad envelope with its maximum at a frequency of about 3\,mHz. The center and the shape of the envelope is defined by the excitation and damping where the later can be assumed to be Gaussian (see previous section). \citet{bro91} and latter on \citet{kje95b} have shown that the frequency of maximum oscillation power, \fm , of p-mode oscillations scales to good approximation with the acoustic cutoff frequency, which sets limits on the maximum frequency for acoustic oscillations. They predict \fm\ by scaling from the Sun as

\begin{equation}
\nu_\mathrm{max} = (M/M\sun ) \cdot (R/R\sun )^{-2} \cdot (T_\mathrm{eff}/5777\,K)^{-1/2} \cdot 3050\,\mu \mathrm{Hz}.
\label{eq:numax}
\end{equation}

It has been shown that this simple scaling relation gives very good estimates for the frequency of maximum oscillation power for less evolved stars \citep[e.g.][]{bed03}, but it cannot a priori be assumed that it holds also for stars of the giant branch. \citet{ste08}, however, have demonstrated for a number of bright red giants observed by the star tracker of the WIRE satellite that the measured \fm\ is in reasonable agreement with the values that can be expected from their fundamental parameters. In their analysis they use the luminosities and effective temperatures determined from Hipparcos parallaxes and infrared photometry, respectively, to substitute the radius in Eq.\,\ref{eq:numax} according to $L \propto R^2 \cdot T^4$.

\begin{table*}[t]
\begin{center}
\caption{Mass and radius of stars used to test our asteroseismic mass and radius determination approach. Mass and radius as taken from the literature are given in brackets.
\label{tab:Tstars}}

\begin{tabular}{lccccccclr}
\hline
\hline
\noalign{\smallskip}
&&$\nu_\mathrm{max}$&$\Delta \nu$&T$_\mathrm{eff}$&&R &M & (R & M)\\ 
&&\multicolumn{2}{c}{[$\mu$Hz]}&[K]&&\multicolumn{4}{c}{- - - - - - solar units - - - - - -}\\
\noalign{\smallskip}
\hline
\noalign{\smallskip}
Arcturus    	&&3.47$\pm$0.03	&0.825$\pm$0.05	&4290$\pm$30		&&27.9$\pm$3.4		&0.8$\pm$0.2&(25.4$\pm$0.3	&0.7-2.0)\\
HD181907	&&29.1$\pm$0.6	&3.47$\pm$0.12	&4760$\pm$65		&&13.1$\pm$1.0		&1.5$\pm$0.23&(12.3$\pm$0.6	&1.5-2.0)\\
$\beta$ Oph 	&&46.0$\pm$2.5	&4.1$\pm$0.2		&4470$\pm$100	&&14.9$\pm$1.7 		&3.1$\pm$0.8 &(12.2$\pm$0.8 	&$\leq$3.2)\\
$\epsilon$ Oph	&&53.5$\pm$2		&5.2$\pm$0.1 		&4877$\pm$100	&&10.7$\pm$0.6		&1.8$\pm$0.3&(10.4$\pm$0.5  	&2.02)\\
$\xi$ Hya 		&&92.3$\pm$3		&7.0$\pm0.2$		&5010$\pm$50		&&10.2$\pm$0.7 		&2.9$\pm$0.4&(10.4$\pm$0.5	 	&2.93-3.15)\\ 
M67 13		&&208.9$\pm$4	&15.9$\pm$0.2		&4966$\pm$50		&&4.6$\pm$0.1			&1.33$\pm$0.1&(4.3$\pm$0.3		&1.35)\\
$\nu$ Ind 		&&313$\pm$10		&24.25$\pm$0.25	&5300$\pm$100 	&&3.04$\pm$0.12		&0.91$\pm$0.1&(2.96$\pm$0.13 	&0.81-0.89)\\
$\zeta$ Her A	&&700$\pm$50		&43$\pm$0.5		&5825$\pm$50		&&2.3$\pm$0.17 		&1.2$\pm$0.26&(2.51$\pm$0.11 	&1.3-1.5)\\ 
$\beta$ Hyi	&&1020$\pm$50	&56.2$\pm$0.2		&5872$\pm$44		&&1.94$\pm$0.1 		&1.28$\pm$0.2&(1.81$\pm$0.02 	&1.0-1.2)\\
HD49933		&&1657$\pm$28	&85.2$\pm$0.5		&6500$\pm$75		&&1.44$\pm$0.03		&1.21$\pm$0.07&(1.46$\pm$0.05	&1.325)\\
$\mu$ Ara	&&1900$\pm$50	&90$\pm$1		&5813$\pm$40 	&&1.40$\pm$0.05 		&1.23$\pm$0.1&(1.36$\pm$0.03 	&1.1-1.3)\\
$\alpha$ Cen A&&2410$\pm$130	&106$\pm$0.2		&5770$\pm$50 	&&1.28$\pm$0.07 		&1.30$\pm$0.2&(1.26$\pm$0.03 	&1.124$\pm$0.008)\\
$\alpha$ Cen B&&4090$\pm$170	&161.4$\pm$0.06	&5300$\pm$50		&&0.90$\pm$0.04 		&1.04$\pm$0.13&(0.85$\pm$0.02 	&0.934$\pm$0.007)\\
\hline
\end{tabular}
\end{center}
\end{table*}

Our sample of red giants is far too faint to measure parallaxes. But the CoRoT observations are significantly better than the WIRE observations in terms of duration and precision, which enables us to also extract the large frequency separation. 
For the Sun, \citet{tou92} have determined a large frequency separation of about 134.92\mh\ 
at the radial order where the maximum oscillation power is seen ($n$ = 21). The large frequency separation reflects essentially the global properties of the star and is believed to scale with the dynamical time scale and therefore with the square root of the mean density. \citet{kje95b} predicted \dnu\ by scaling from the Sun as
 
\begin{equation}
\Delta \nu = (M/M\sun )^{1/2} \cdot (R/R\sun )^{-3/2} \cdot 134.92\,\mu \mathrm{Hz}.
\label{eq:dnu}
\end{equation}
Knowing $\nu_\mathrm{max}$ and $\Delta \nu$ (and T$_\mathrm{eff}$), it is now easy to derive the stellar mass and radius: 
\begin{equation}
R/R\sun =  (\nu_\mathrm{max}/\nu_\mathrm{max,\, \sun}) \cdot  (\Delta \nu / \Delta \nu \sun )^{-2} \cdot (T_\mathrm{eff}/5777\,K)^{1/2}
\label{eq:r}
\end{equation}
\begin{equation}
M/M\sun = (R/R\sun )^3 \cdot (\Delta \nu /\Delta \nu \sun )^2 .
\label{eq:m}
\end{equation}
We have tested this method for a number of well-known solar-type pulsators and compare in Table\,\ref{tab:Tstars} the seismic masses and radii with independent measurements given in the literature. The specific values for the latter are taken from
\begin{itemize}
\item Arcturus: \fm\ and \dnu\ are given by \citet{tar07} and \citet{ret03}, respectively. The radius is based on the Hipparcos parallax \citep{lee07} and interferometric measurements \citep{lac08}. The mass range is estimated from the average surface gravity listed in the VizieR\footnote{http://vizier.u-strasbg.fr/viz-bin/VizieR} database (\lg\ = 1.72$\pm$0.2) and the interferometric radius.
\item HD181907: We determine \fm\ from the CoRoT observations published in \citet{car09}, who also derived \dnu . The mass is estimated from a comparison between the star's position in the H-R diagram and metal-poor evolutionary tracks.
\item $\beta$ Oph: \fm\ and \dnu\ were determined from unpublished MOST photometry. The radius is based on the Hipparcos parallax \citep{lee07} and interferometric measurements \citep{ric05}. An upper limit for the mass is estimated from the average surface gravity (\lg\ = 2.42$\pm$0.3) taken from the VizieR database and the interferometric radius.
\item $\epsilon$ Oph:  \fm\ and \dnu\ are given by \citet{kal08}, who determined the mass from a detailed comparison of observed and model frequencies. The radius is based on the Hipparcos parallax \citep{lee07} and interferometric measurements \citep{ric05}.
\item $\xi$ Hya: \fm , \dnu , and the mass are taken from \citet{fra02} ,where they estimate the mass from a comparison between the star's position in the H-R diagram and solar-calibrated evolutionary tracks. We derive \fm\ from a weighted average of the published frequencies. 
\item M67 13: \fm\ and \dnu\ are determined from the photometric time series kindly provided by D. Stello. The mass is estimated from isochrone fits to the color-magnitude diagram of M67 \citep{ste07}.
\item $\nu$ Ind: \dnu\ is given by \citet{car07}. \fm\ is taken from \citet{bed06}, who also provide a mass range based on a comparison between the star's position in the H-R diagram and evolutionary tracks.
\item $\zeta$ Her A: \fm , \dnu , and the mass are taken from \citet{mar01}. 
\item $\beta$ Hyi: \fm\ and \dnu\ are given by \citet{kje05} and \citet{bed01}, respectively. The radius is based on interferometric measurements from \citet{nor07}, who also provide a summary of non-seismically determined values for the mass.
\item HD\,49933:  \fm\ and \dnu\ are given by \citet{kal09}, who determined the mass from a detailed comparison of observed and model frequencies. 
\item $\mu$ Ara: \fm\ and \dnu\ were extracted from \citet{bou05}. A non-seismic estimate for the mass can be found in \citet{baz05}.
\item $\alpha$ Cen A \& B: \fm\ and \dnu\ are taken from \citet{kje05}. Masses were determined by \citet{gue00} using Hipparcos parallaxes and the binary mass ratio.
\end{itemize}

   \begin{figure*}[t]
   \centering
      \includegraphics[width=0.97\textwidth]{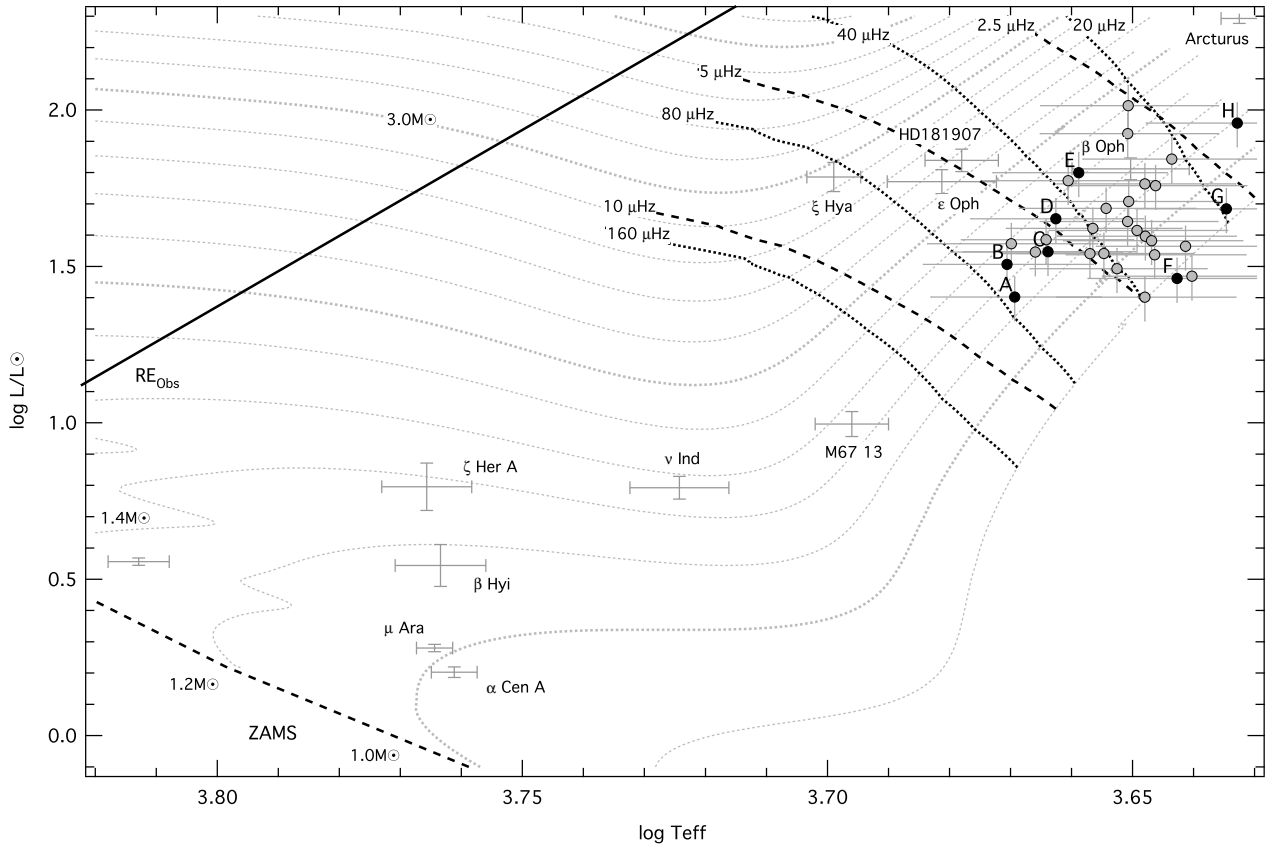}
      \caption{Theoretical H-R diagram showing the location of the stars used to test our asteroseismic mass and radius determination approach. Grey-filled dots (total sample) and black dots (stars presented in Fig.\,\ref{Fig:Fourier}) indicate the analyzed CoRoT pulsating red giants where the actual position in the HR-diagram is based on a comparison with evolutionary tracks. The errors bars correspond to the approximate uncertainties of our method and are significantly larger than the observationally based errors. Dashed black lines indicate isopleths for $\Delta \nu$ = 2.5, 5, and 10\mh\ and dotted black lines show isopleths for $\nu_\mathrm{max}$ = 20, 40, 80, and 160\mh\ where we linearly interpolate for points between the evolutionary tracks.} 
         \label{Fig:HRD}
   \end{figure*}

If not explicitly mentioned the radius is determined according to $L \propto R^2 \cdot T_\mathrm{eff}^4$ with the effective temperature and luminosity taken from the listed reference (or references therein).
Although \fm\ is in some cases only a rough estimate from a published power spectrum, our approach yields quite accurate masses and radii in a large portion of the H-R diagram. 
For faint stars it often turns out that no or only poor estimates for the effective temperature are available. This is, however, not very critical for stars on the giant branch, as \fm\ depends only on the square root of $T_\mathrm{eff}$, and pulsating red giants are expected to populate only a relatively narrow temperature range ($\sim$4200 to 5300\,K). It should therefore be possible to get a reasonable asteroseismic mass and radius for red giants even if an accurate $T_\mathrm{eff}$ is not available. This is shown for the red giants in Table\,\ref{tab:Tstars} (Arcturus to $\xi$ Hya), where we ignore the known values for $T_\mathrm{eff}$ and fix the temperature to a typical value of 4750\,K.

\begin{table*}[t]
\begin{center}
\caption{Summary for the analysed sample of red gaints. V magnitudes are taken from the \textit{EXODAT} database \citep{del06}. \fm\ and \dnu\ are obtained from the photometric time series. Fundamental parameters are based on scaling relations for \fm\ and \dnu\ and interpolation in a grid of solar-calibrated RGB models. The errors are based \emph{only} on observational uncertainties. 
\label{tab:Cstars}}
\begin{tabular}{cccccccclr}
\hline
\hline
\noalign{\smallskip}
&CoRoT-ID&2MASS&V [mag] & $\nu_\mathrm{max}$ [$\mu$Hz] & $\Delta \nu$ [$\mu$Hz] & R/R\sun & M/M\sun & (T$_\mathrm{eff}$  &  L/L\sun )\\
\noalign{\smallskip}
\hline
\noalign{\smallskip}
A  & 101113062& 19263135-0002147	& 13.46 &74.32$\pm$0.81&  7.20$\pm$0.04&   7.69$\pm$0.11&  1.30$\pm$0.05&  (4670$\pm$10&  25$\pm$1) \\
B  & 101251252& 19272346+0119408	& 13.72 &67.74$\pm$0.92&  6.50$\pm$0.03&   8.62$\pm$0.15&  1.49$\pm$0.07&  (4684$\pm$14&  32$\pm$1) \\
C  & 101362522& 19281037+0049531	& 13.90 &53.51$\pm$1.28&  5.53$\pm$0.04&   9.32$\pm$0.26&  1.36$\pm$0.11&  (4612$\pm$21&  35$\pm$3) \\
D  & 101034881& 19255782+0056022	& 13.50 &45.59$\pm$0.90&  4.79$\pm$0.02&  10.58$\pm$0.23&  1.49$\pm$0.09&  (4598$\pm$18&  45$\pm$3) \\
E  & 101197556& 19270316-0032250	& 13.35 &34.60$\pm$0.69&  3.79$\pm$0.04&  12.75$\pm$0.40&  1.64$\pm$0.13&  (4558$\pm$23&  63$\pm$5) \\
F  & 100838545& 19245182+0129416	& 12.76 &31.03$\pm$0.65&  4.15$\pm$0.05&   9.38$\pm$0.29&  0.78$\pm$0.06&  (4395$\pm$16&  29$\pm$2) \\
G  & 101649216& 19301425-0007224	& 12.46 &19.30$\pm$0.37&  2.82$\pm$0.11&  12.48$\pm$1.00&  0.85$\pm$0.14&  (4311$\pm$26&  48$\pm$9) \\
H  & 101378942& 19281895+0105256	& 12.76 &14.03$\pm$0.44&  2.05$\pm$0.03&  17.25$\pm$0.69&  1.18$\pm$0.13&  (4294$\pm$22&  91$\pm$9) \\
\noalign{\smallskip}
   & 101290847& 19273800+0112570	& 13.20 &61.13$\pm$0.60&  5.93$\pm$0.11&   9.33$\pm$0.35&  1.57$\pm$0.12&  (4676$\pm$21&  37$\pm$3) \\
   & 101509360& 19291255-0011005	& 13.06 &56.77$\pm$0.66&  5.74$\pm$0.03&   9.22$\pm$0.16&  1.42$\pm$0.06&  (4634$\pm$12&  35$\pm$2) \\
   & 101081290& 19261943-0032595	& 12.73 &51.59$\pm$0.42&  5.32$\pm$0.06&   9.72$\pm$0.20&  1.43$\pm$0.07&  (4615$\pm$11&  38$\pm$2) \\
   & 101041814& 19260095+0131378	& 13.51 &43.40$\pm$0.57&  4.90$\pm$0.06&   9.55$\pm$0.24&  1.15$\pm$0.07&  (4539$\pm$12&  35$\pm$2) \\
   & 100886873& 19250710+0106045	& 13.57 &40.89$\pm$0.75&  4.73$\pm$0.04&   9.65$\pm$0.24&  1.10$\pm$0.07&  (4515$\pm$15&  35$\pm$2) \\
   & 100483847& 19224577+0131126	& 12.70 &39.15$\pm$1.07&  4.44$\pm$0.03&  10.50$\pm$0.30&  1.25$\pm$0.10&  (4534$\pm$19&  42$\pm$3) \\
   & 101232297& 19271621-0015395	& 12.48 &40.02$\pm$0.91&  4.78$\pm$0.03&   9.22$\pm$0.24&  0.98$\pm$0.07&  (4493$\pm$19&  31$\pm$2) \\
   & 100974118& 19253501+0022085	& 13.74 &39.03$\pm$0.66&  4.91$\pm$0.06&   8.48$\pm$0.25&  0.81$\pm$0.06&  (4446$\pm$15&  25$\pm$2) \\
   & 100716817& 19241108+0138078	& 13.86 &37.40$\pm$0.77&  4.02$\pm$0.03&  12.28$\pm$0.31&  1.65$\pm$0.11&  (4577$\pm$22&  59$\pm$4) \\
   & 101218811& 19271113+0152095	& 13.46 &34.53$\pm$1.10&  3.99$\pm$0.04&  11.42$\pm$0.44&  1.30$\pm$0.14&  (4512$\pm$25&  49$\pm$5) \\
   & 101242228& 19271998+0045148	& 13.10 &31.17$\pm$0.70&  3.92$\pm$0.08&  10.62$\pm$0.47&  1.01$\pm$0.10&  (4445$\pm$23&  40$\pm$4) \\
   & 101136306& 19264008+0127108	& 12.61 &32.22$\pm$0.58&  3.91$\pm$0.04&  11.06$\pm$0.31&  1.14$\pm$0.08&  (4475$\pm$15&  44$\pm$3) \\
   & 101058180& 19260841+0151442	& 12.61 &31.84$\pm$0.77&  4.08$\pm$0.05&   9.98$\pm$0.35&  0.91$\pm$0.08&  (4430$\pm$21&  34$\pm$3) \\
   & 100908597& 19251396+0034026	& 13.95 &31.76$\pm$1.10&  3.93$\pm$0.04&  10.78$\pm$0.43&  1.06$\pm$0.12&  (4459$\pm$26&  41$\pm$4) \\
   & 101262795& 19272783+0100075	& 14.16 &30.75$\pm$0.81&  3.92$\pm$0.07&  10.49$\pm$0.47&  0.97$\pm$0.10&  (4435$\pm$24&  38$\pm$4) \\
   & 101044584& 19260224-0019015	& 13.00 &28.53$\pm$0.98&  3.95$\pm$0.08&   9.49$\pm$0.51&  0.73$\pm$0.09&  (4368$\pm$24&  29$\pm$4) \\
   & 101513442& 19291426+0009508	& 14.32 &30.06$\pm$0.80&  3.64$\pm$0.05&  11.91$\pm$0.46&  1.23$\pm$0.12&  (4473$\pm$21&  51$\pm$5) \\
   & 100855073& 19245704+0130376	& 14.13 &26.37$\pm$0.52&  3.28$\pm$0.02&  12.87$\pm$0.32&  1.26$\pm$0.08&  (4446$\pm$14&  58$\pm$4) \\
   & 101150795& 19264557+0051573	& 14.11 &25.05$\pm$0.61&  3.19$\pm$0.07&  12.88$\pm$0.69&  1.19$\pm$0.15&  (4428$\pm$23&  57$\pm$7) \\
   & 101654204& 19301756-0014319	& 13.59 &26.71$\pm$0.90&  3.63$\pm$0.03&  10.54$\pm$0.40&  0.85$\pm$0.09&  (4378$\pm$25&  37$\pm$4) \\
   & 100998571& 19254285-0012038	& 13.70 &24.29$\pm$0.52&  2.89$\pm$0.04&  15.27$\pm$0.56&  1.64$\pm$0.14&  (4475$\pm$23&  84$\pm$8) \\
   & 101449976& 19284824-0007431	& 12.38 &21.50$\pm$0.56&  2.79$\pm$0.02&  14.38$\pm$0.43&  1.27$\pm$0.11&  (4401$\pm$17&  70$\pm$5) \\
   & 101029979& 19255560+0122554	& 14.64 &21.85$\pm$1.05&  2.60$\pm$0.03&  16.95$\pm$0.89&  1.81$\pm$0.27&  (4474$\pm$44&  100$\pm$15) \\
\noalign{\smallskip}
\hline
\end{tabular}
\end{center}
\end{table*}

For our CoRoT sample of red giants we first estimate the effective temperature from 2MASS photometric colors and color-temperature calibrations \citep{mas06,gon09}. Unfortunately, both calibrations result in temperatures which are systematically too cool. The average temperature resulting from the \citet{mas06} calibration is about 3820\,K, which would make our sample of red giants either extremely metal rich or would put all the stars high up on the giant branch. Both explanations are not very plausible. We expect the discrepancy to be due to severe reddening which is difficult to estimate for such faint (and therefore distant) stars. The situation is sligthly better for the \citet{gon09} calibration, but we decided to ignore the effective temperatures determined from color-temperature calibrations.

Instead, we use a different approach. We calculate an initial guess for the mass and radius from Eqs.\,\ref{eq:r} and \ref{eq:m} by fixing the effective temperature to a typical value of 4750\,K. In a next step we compare the initial mass and radius to those of a grid of solar-calibrated red giant models. Interpolation in the grid gives a better estimate for the temperature, which is used as a new input for Eq.\,\ref{eq:r}. After about three iterations the procedure converges to a certain location in the H-R diagram where the final locus in the H-R diagram is independent from the starting value for T$_\mathrm{eff}$ as long as the initial value is kept within the temperature range of our model grid ($\sim$4000 - 5500\,K).

The resulting fundamental parameters and their uncertainties are given in Table\,\ref{tab:Cstars}. The errors are based only on the uncertainties for \fm\ and \dnu\ and range from 1.5 to 8.2\% and 3.9 to 17\% for the radius and mass, respectively.

The red-giant models used to estimate the effective temperatures fall along evolutionary tracks computed with the Yale Stellar Evolution Code YREC \citep{gue92,dem07}. The evolutionary tracks were computed for an initial helium and metal mass fraction (Y, Z) = (0.28, 0.02) with the mixing-length parameter $\alpha$ = 1.8 set to approximately meet the Sun's position in the H-R diagram with a one-solar-mass model at roughly the solar age. Note, although the models are not exactly calibrated to the Sun, we refer to them as \textit{solar-calibrated} in the following. A more detailed description of the used model physics can be found in \citet{kal08} or \citet{kal09t} and references therein. 

In Fig.\,\ref{Fig:HRD} we show our sample of red giants in the H-R diagram along with the test stars (Table\,\ref{tab:Tstars}) and contours of constant \fm\ and \dnu . We whish to emphasize that the actual locations of the red giants are specific to the model grid used to estimate their effective temperatures. We assume that all red giants are comparable to the Sun in terms of their initial chemical composition and mixing-length parameter. This might be correct for some of them but not for others. The effective temperature and luminosity of a model with a given mass and radius do not only depend on the model's age but also for instance on the initial chemical composition of the model. Metal-poor models, e.g., are shifted towards higher surface temperatures and luminosities compared to solar-abundant models. 
On the other hand, low and intermediate mass red giants contract rapidly after they have reached the tip of the red giant branch (RGB) to settle on the He-core burning main sequence at a somewhat higher temperature before they start to climb the asymptotic giant branch (ABG). 
In other words, a red giant with a given mass and radius is located at different positions in the H-R diagram depending on, e.g., the chemical composition and/or the evolutionary stage.

\begin{figure}[t]
   \centering
      \includegraphics[width=0.5\textwidth]{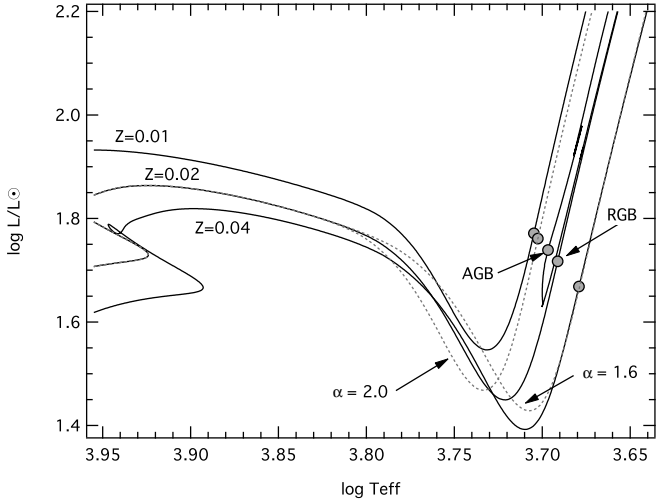}
      \caption{YREC evolutionary tracks for 2.5\,M\sun\ models with different chemical compositions (Y, Z) = (0.25, 0.01), (0.28, 0.02), and (0.32, 0.04) and mixing-length parameters $\alpha$ = 1.6, 1.8, and 2.0. For the solar-calibrated track also the AGB part of the track is shown. The large dots indicate models with 10\,R\sun .}
         \label{Fig:HRDdiff}
   \end{figure}

To illustrate this ambiguity we compare stellar models for a given mass and radius, but with different initial chemical composition and during different evolutionary stages. The result is shown in Fig.\,\ref{Fig:HRDdiff} where we illustrate that 2.5\,M\sun\ RGB models with 10\,R\sun\ with a initial chemical composition of (Y, Z) = (0.25, 0.01) and (0.32, 0.04) are about 155K hotter and 13\% more luminous and about 135K cooler and 11\% less luminous, respectively, than a solar-calibrated RGB model with the same mass and radius. Similar results can be expected for different mixing-length parameters. Whereas the parameterization of convection has only small effects on the surface properties of a star during early evolution, different mixing-length parameters result in quite different evolutionary tracks when the star ascends the giant branch. This is because the mixing-length parameter sets the temperature gradient in the convective regions and thus controls how efficiently energy can be deduced from the interior. Consequently, the mixing-length parameter defines at what stage the star starts to climb the giant branch during the hydrogen shell burning phase. 
A slightly smaller effect can be expected for the different evolutionary stages. Our 2.5\,M\sun\ and 10\,R\sun\ is about 60K hotter and 5\% more luminous in its AGB phase than the corresponding RGB model. Although most of our sample red giants are expected to have a mass below 2.5\,M\sun , we use here the more massive models. 
This is because YREC, just like other stellar evolution codes, is not able to follow the explosive He flash in low mass stars, and we currently have no AGB starting models available. We can follow the He core ignition only for higher mass models and evolve models on the AGB. We do not expect the effect to be significantly different for lower mass models, however.

We have demonstrated in Fig.\,\ref{Fig:HRDdiff} that an unknown chemical composition, mixing-length parameter, and evolutionary stage adds a certain amount of uncertainty when we determine a star's position in the H-R diagram from its mass and radius. This ambiguity will add additional uncertainties on the asteroseismic masses and radii. To quantify this effect we have computed additional sets of evolutionary tracks. For the sake of simplicity and as we expect the uncertainty to be smaller for an unknown evolutionary stage than for an unknown initial chemical composition we concentrate here on models with different initial chemical compositions, namely (Y, Z) = (0.25, 0.01) and (0.32, 0.04). In Table\,\ref{tab:Comp} we compare the fundamental parameters as they result from using the different model grids to estimate the effective temperature for Eq.\,\ref{eq:r}. We list here only three stars which are, however, selected to cover the range of interest in the H-R diagram. 
The spread in chemical composition adds an additional uncertainty of not more than about 1.8 and 5.5\% on the radius and mass, respectively, which is smaller or at most comparable to the observationally based errors. The situation is different for the effective temperature and luminosity where the additional uncertainties are significantly larger than the observational errors. To account for this we have extended the error bars in Fig.\,\ref{Fig:HRD} to $\pm$150\,K and $\pm$17\% for the effective temperature and luminosity, respectively. Note that these are still quite accurate fundamental parameters for such faint stars.

\begin{figure}[t]
   \centering
      \includegraphics[width=0.5\textwidth]{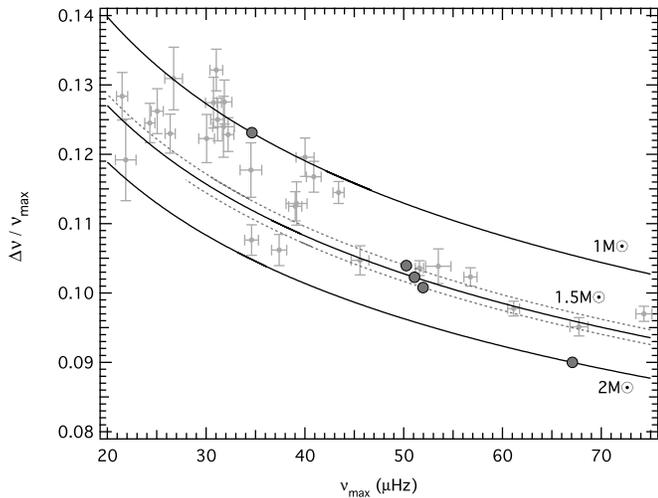}
      \caption{Diagnostic asteroseismic diagram for stars on the red-giant branch. Solid lines correspond to solar-calibrated evolutionary tracks. The dotted lines indicate models with 1.5\,M\sun , but with different initial chemical compositions (the upper and lower line corresponds to the metal-poor and metal-rich models, respectively). The large dots indicate 10\,R\sun\ models, and the light-grey dots are the CoRoT red giants listed in Table \ref{tab:Cstars}.  }
         \label{Fig:DiaDia}
   \end{figure}

From Eqs.\,\ref{eq:numax} and \ref{eq:dnu} and the isopleths for \fm\ and \dnu\ in Fig.\,\ref{Fig:HRD} it is obvious that the two parameters are correlated to some extent. Both parameters depend on the stellar mass and radius. \citet{ste09} indeed found a tight relation between \fm\ and \dnu\ for main-sequence and red-giant stars, which was confirmed by \citet{hek09} for CoRoT red giants. Although the relation seems to be very tight, no \textit{exact} relation can be explained from a theoretical point of view. It even turns out that the relation between \fm\ and \dnu\ strongly depends on the stellar mass. This can be seen from Fig.\,\ref{Fig:DiaDia}, where we compare the ratio between \dnu\ and \fm\ as a function of \fm\ for models of different evolutionary tracks with the sample of red giants. The ordinate basically represents the inverse radial order where the maximum oscillation power is seen \citep{kje95b}.
We introduce this diagram as a sort of diagnostic asteroseismic diagram similar to a diagram which is usually used to estimate the mass and central hydrogen abundance for solar-type pulsators close to or on the main sequence from the measured large and small frequency separations \citep[see e.g., ][]{rox03}. Our diagnostic asteroseismic diagram is particularly useful on the giant branch and relies on an observational quantity, namely \fm , which is easier to determine than the small frequency separation. Other parameters like the helium core abundance can easily be added to this diagram.
The diagram also demonstrates the relative robustness of our method to determine an asteroseismic mass and radius. Models of a given mass and radius (large dots) are only slightly shifted in the diagram if e.g. their initial chemical composition is significantly changed.

We have illustrated to some extent what uncertainty can be expected for a red giant's position in the H-R diagram if the star's initial chemical composition and/or mixing-length parameter and/or evolutionary stage is unknown. But there are also other effects in stellar evolution which carry a similar type of uncertainty. Examples are the overshoot parameter or a better description of convection than the MLT which change the L-M-R-T$_\mathrm{eff}$ relations. These effects are difficult to estimate and are not on the scope of the current analysis.

\begin{table}[t]
\begin{center}
\caption{Fundamental parameters for the stars A, E, and H as they follow from using a metal-poor (MP), solar-calibrated (SC), or metal-rich (MR) grid to estimate the effective temperature. Also given are standard deviations in absolute units and percent. 
\label{tab:Comp}}
\begin{tabular}{cccclr}
\hline
\hline
\noalign{\smallskip}
ID&Grid& R/R\sun & M/M\sun & T$_\mathrm{eff} [K]$  &  L/L\sun \\
\noalign{\smallskip}
\hline
\noalign{\smallskip}
     &MP&  7.82$\pm$0.12&  1.36$\pm$0.05&  4835$\pm$11&  30$\pm$1 \\
A   &SC&   7.69$\pm$0.11&  1.30$\pm$0.05&  4670$\pm$10&  25$\pm$1 \\
     &MR&  7.57$\pm$0.12&  1.24$\pm$0.05&  4528$\pm$10&  22$\pm$1 \\
\noalign{\smallskip}
\hline
\noalign{\smallskip}
\multicolumn{2}{r}{$\sigma$}&  0.12(1.6\%)&  0.06(4.6\%)&  153(3.3\%)&  4(15\%) \\
\noalign{\smallskip}
\hline
\noalign{\smallskip}
    &MP&  12.98$\pm$0.39&  1.73$\pm$0.13&  4725$\pm$25&  75$\pm$6 \\
E  &SC&  12.75$\pm$0.40&  1.64$\pm$0.13&  4558$\pm$23&  63$\pm$5 \\
    &MR&  12.56$\pm$0.38&  1.56$\pm$0.12&  4417$\pm$19&  54$\pm$4 \\
\noalign{\smallskip}
\hline
\noalign{\smallskip}
\multicolumn{2}{r}{$\sigma$}&  0.21(1.6\%)&  0.09(5.5\%)&  154(3.3\%)&  11(17\%) \\
\noalign{\smallskip}
\hline
\noalign{\smallskip}
    &MP&  17.59$\pm$0.71&  1.25$\pm$0.13&  4462$\pm$46&  110$\pm$10 \\
H  &SC&  17.25$\pm$0.69&  1.18$\pm$0.13&  4294$\pm$22&  91$\pm$9 \\
    &MR&  16.98$\pm$0.71&  1.13$\pm$0.12&  4160$\pm$28&  78$\pm$9 \\
\noalign{\smallskip}
\hline
\noalign{\smallskip}
\multicolumn{2}{r}{$\sigma$}&  0.31(1.8\%)&  0.06(5.0\%)&  151(3.5\%)&  16(17\%) \\
\noalign{\smallskip}
\hline
\noalign{\smallskip}

\end{tabular}
\end{center}
\end{table}

\section{Conclusions and prospects}
We have shown that global properties of solar-type pulsations can be used to derive estimates for the stellar mass and radius by employing well-established and often used scaling relations. We have tested this approach on various prominent solar-type pulsators and applied it to a first sample of red giant pulsators observed by CoRoT. Despite the mentioned approximations the derived fundamental parameters can serve to constrain the starting values for a more detailed analysis. 

We note that we do not stop at this point. In a next step we will use the integral of the Gaussian part of our global power density model to deduce the total spectral power of solar-type pulsations, which we expect to scale with the luminosity-mass ratio. This, however, needs extensive calibration for solar-type pulsators with independently determined fundamental parameters, which we are currently carrying out. We believe that we can, for the first time, derive all basic fundamental parameters (mass, radius, luminosity, and consequently also the effective temperature) of a solar-type pulsator by simply measuring global properties of its oscillations in the power density spectrum. This will have influence on various astrophysical applications. One can use it as a distance indicator, or one can study the behavior of convective time scales of stars as a function of their position in the H-R diagram. By comparing the individual pulsation frequencies with theoretical eigenfrequencies it should be possible to investigate the parameterization of convective models (e.g, the mixing-length parameter) in a region of the H-R diagram where stars are very sensitive to these parameters.

Finally, we want to mention that we will apply our asteroseismic fundamental parameter determination to all pulsating red giants observed by CoRoT, and also we plan to arrange an online database for them.

\begin{acknowledgements}
TK, MG, and WWW are supported by the Austrian Research Promotion Agency (FFG), and the Austrian Science Fund (FWF P17580). TK is also supported by the Canadian Space Agency. The research leading to these results has received funding from the Research Council of K.U.Leuven under grant agreement GOA/2008/04 and from the Belgian PRODEX Office under contract C90309: CoRoT Data Exploitation.
FC is a postdoctoral fellow of the Fund for Scientific Research, Flanders. APH acknowledges the support grant 50OW0204 from the Deutsches Zentrum f\"ur Luft- und Raumfahrt e. V. (DLR). SH acknowledges financial support from the Belgian Federal Science Policy (ref: MO/33/018). CC is supported partially be a CITA national fellowship. Furthermore, it is a pleasure to thank D. Stello (University of Sydney) for providing us with the photometric data of M67. We thank the MOST Science Team for letting us use the unpublished photometry of $\beta$ Oph. Finally, we thank the anonymous referee for helping us to improve the manuscript. 
\end{acknowledgements}

\end{document}